# Expansion of Solar Coronal Hot Electrons in an Inhomogeneous Magnetic Field: 1-D PIC Simulation

Jicheng Sun[1], Xinliang Gao[1*], Yangguang Ke[1], Quanming Lu[1#], Xueyi Wang[2], and Shui Wang[1]

[1]CAS Key Laboratory of Geospace Environment, Department of Geophysics and Planetary Science, University of Science and Technology of China, Hefei, China

[2]Physics Department, Auburn University, Auburn, Alabama, USA

*Email: gaoxl@mail.ustc.edu.cn

#Email: qmlu@ustc.edu.cn

# Abstract

The expansion of hot electrons in flaring magnetic loops is crucial to understanding the dynamics of solar flares. In this paper we investigate, for the first time, the transport of hot electrons in a magnetic mirror field based on a 1-D particle-in-cell (PIC) simulation. The hot electrons with small pitch angle transport into the cold plasma, which leads to the generation of Langmuir waves in the cold plasma and ion acoustic waves in the hot plasma. The large pitch angle electrons can be confined by the magnetic mirror, resulting in the different evolution time scale between electron parallel and perpendicular temperature. This will cause the formation of electron temperature anisotropy, which can generate the whistler waves near the interface between hot electrons and cold electrons. The whistler waves can scatter the large pitch angle electrons to smaller value through the cyclotron resonance, leading to electrons escaping from the hot region. These results indicate that the whistler waves may play an important role in the transport of electrons in flaring magnetic loops. The findings from this study provide some new insights to understand the electron dynamics of solar flares.

# I. Introduction

Thermal conduction is a fundamental process occurring in space and astrophysical plasma, which is considered to play a significant role in solar flares and the intracluster medium of galaxy clusters. During solar flares, a significant amount of magnetic energy is converted into energetic electrons through magnetic reconnection (e.g., Fu et al. 2006; Huang et al. 2010; Dahlin et al. 2014; Wang et al. 2016; Li et al. 2017; Lu et al. 2018). Generally, these energetic electrons can emit hard X-ray (HXR) through interaction with background plasma both in the corona (Masuda et al. 1994; Krucker et al. 2007, 2008) and chromosphere (e.g., Hoyng et al. 1981). How these energetic electrons transport from the coronal acceleration site to the chromosphere plays a crucial role in understanding the dynamics of flares, since transport effects can affect the interpretation of flare models (Battaglia & Benz 2006). Krucker et al. (2007) made a study of coronal HXR emissions, finding that the HXR emissions can last up to several minutes in the corona. This is two orders of magnitude longer than the free-streaming transit time of electrons through the source region (Masuda et al. 1994; Krucker et al. 2007, 2010), which suggests that HXR-producing electrons should be trapped in the source region of the corona. The magnetic mirror (Simões & Kontar 2013) and thermal fronts (Rust et al. 1985; Batchelor et al. 1985) provide a potential mechanism for the confinement of hot electrons in flaring magnetic loops. Using the test particle method, Varady et al. (2014) have investigated the transport of energetic electrons including the influence of the magnetic mirror and the electric field. Regarding to the intracluster medium, it is found that a large fraction of the energy injection by a central jetted active galactic nucleus is thermalized in the intracluster medium (Churazov et al. 2000, 2002;

Reynolds et al. 2002). Understanding these astrophysical phenomena also requires the knowledge of electron transport processes.

The transport of hot electrons in a uniform magnetized plasma has been thoroughly studied using theoretical analysis (Brown et al. 1979; Smith & Lillequist 1979) and numerical simulations (Ishigura et al. 1985; Arber & Melnikov 2009; Li et al. 2012; Karlický 2015; Sun et al. 2019). Theoretical analysis shows that hot electrons transport along the magnetic field into a cool surrounding plasma, leading to a return current from the background cold electrons. Then, the return current can generate ion acoustic turbulence which efficiently scatters the electrons, inhibiting the transport of electron heat flux and forming thermal fronts (Manheimer 1977; Brown et al. 1979). McKean et al. (1990) used a PIC simulation to study this process and claimed that a thermal front is not established for a plasma inhomogeneous in electron temperature and homogeneous in density and ion temperature. However, recent simulations (Arber & Melnikov 2009) have indicated that the absence of thermal fronts is due to the restricted size of the initial hot-electron region. The thermal fronts can exist when larger hot region is used. Sun et al. (2019) made an extended study of thermal fronts using a 1D PIC simulations. They found that the thermal front will experience the dissipation and reformation process during the expansion of hot electrons.

All previous PIC simulations of the electron transport in solar flares were performed in a uniform background magnetic field, while flaring magnetic loops are often mirror shaped. Therefore, the self-consistent evolution on the expansion of hot electrons just remains unknown in an inhomogeneous magnetic field, which is critical in understanding the dynamics of solar flares. In this paper we study, for the first time, the expansion of hot electrons in a magnetic mirror field using a 1-D PIC simulation. The structure of this paper is as follows. We first describe the PIC simulation model and initial parameters in section 2, followed by the simulation results in section 3. At last, conclusions and discussion are given in section 4.

## II. Simulation Model

The PIC simulation model is a powerful tool to investigate the plasma process via a self-consistent way (Sun et al. 2016, 2017; Lu et al. 2016; Ke et al. 2017; Gao et al. 2018; Chen et al. 2018). In this paper, a 1-D PIC simulation is used to study the expansion of hot electrons in a magnetic mirror field. The model retains 3-D electromagnetic fields and particle velocities but only 1-D spatial variations in the $z$ direction. The background magnetic field **B₀** lies along the $z$ axis and is defined as

$$B_{0z} = (1 + \xi z^2)B_0 , \qquad (1)$$

where $B_0$ represents the magnetic field at $z = 0$ and $\xi$ is a parameter representing the inhomogeneity of the background magnetic field. The simulation domain is in the ranges $[-L/2, L/2]$. For the initial condition, the plasma consists of three components: cold electrons, cold protons, and hot electrons. The density of each components $n_0$ is uniform. All distribution functions are initially Maxwellian with zero net flow. The hot electrons are distributed in the middle of the computational domain ($-L/20 < z <$ $L/20$), while the cold electrons are located elsewhere. Cold protons and electrons have the same temperature. The temperature of hot electrons is 10 times that of cold electrons.

In this simulation, the magnetic field is normalized to $B_0$, and the temperature is expressed in the units of the cold electron temperature $T_0$. The time and space are normalized to the inverse of electron gyrofrequency $\Omega_e = eB_0/m_e$ and the electron gyroradius $\rho_e = v_{t0}/\Omega_e$ (where $v_{t0}$ is the thermal speed of the cold electrons) at $z = 0$, respectively. For reducing computational cost, the mass ratio of proton to electron is reduced such that $m_i/m_e = 100$. The speed of light $c = 100V_A$ is adopted, where $V_A = B_0/\sqrt{\mu_0 n_0 m_i}$ is the Alfven speed at the coordinate origin. The cold electron plasma beta at the coordinate origin is initially set as $\beta_{ec} = 0.4$, and then the Alfven speed is $V_A = 0.224 v_{t0}$. The time step is set as $\Omega_e \Delta t = 0.001$ such that electron dynamics can be fully resolved. The total length of the computational size is $L = 4000\rho_e$, and the number of grid cells is $N_z = 80000$. The inhomogeneity of the background magnetic field is set as $\xi = 10^{-5}\rho_e^{-2}$. There are on average 1000 macroparticles in every cell for each species. The reflecting boundary conditions are used for particles, and absorbing boundary conditions are assumed for electromagnetic fields (Tao 2014; Lu et al. 2019).

## III. Simulation Results

With the 1-D PIC simulation model and the simulation setup described above, we present the expansion of hot electrons in a magnetic mirror field. Since this simulation model is a symmetric system, it is equivalent to analyze either side of the contact of hot electrons with cool electrons. We choose the right side to examine the expansion of hot electrons without any preference, where hot electrons propagate toward the positive direction.

Figure 1(a) and 1(b) show the temporal and spatial evolution of parallel and perpendicular temperature of electrons, respectively. The temperature anisotropy of electrons $T_{e\perp}/T_{e\|}$ is displayed in Figure 1(c). The parallel and perpendicular temperature of electrons is calculated as

$$T_{e\|} = m_e \langle (v_{ez} - \langle v_{ez} \rangle)^2 \rangle$$

and

$$T_{e\perp} = \frac{m_e}{2} \langle (v_{ex} - \langle v_{ex} \rangle)^2 + (v_{ey} - \langle v_{ey} \rangle)^2 \rangle,$$

where the angle brackets denote an average over particles inside a cell. The method to calculate the temperature has also been used in Lu & Li (2007). Initially, The hot electrons are distributed in $z < 200\rho_e$, and cold electrons are located in $z > 200\rho_e$. There is a steep temperature gradient between the hot and cold electrons. During $\Omega_e t < 40$, the hot electrons with small pitch angle freely expand into the cold plasma, leading to the rapid decrease of parallel temperature in the hot-electron region. According to the previous simulation results in the uniform magnetic field (Li et al. 2012; Sun et al. 2019), the expansion of hot electrons induces a polarized electric field, which will draw a return current from the background cold electrons. This return current can generate ion acoustic waves, evolving into a double layer. For $\Omega_e t > 40$, the parallel temperature of hot electrons slowly drops over time because of the double layer. We can find that the perpendicular temperature also drops in Figure 1(b) but with the longer time scale compared with Figure 1(a). At $\Omega_e t = 60$, the electron parallel temperature is reduced to 60% of its initial temperature, while the perpendicular temperature of electron is almost unchanged. The perpendicular temperature can remain longer time since the magnetic mirror can confine hot electrons with large pitch angle. This will cause the temperature anisotropy of electrons. As shown in Figure 1(c), the electron temperature anisotropy occurs in the hot region at $\Omega_e t = 40$, which can lead to the excitation of whistle waves. We will analysis the generation and evolution of the whistle waves in detail in Figure 3.

Similar to the simulation results in uniform background magnetic field, Langmuir waves and ion acoustic waves can also be excited in non-uniform magnetic field. We examine the evolution of the parallel fluctuating electric fields in Figure 2(a). Figure 2(b) and 2(c) show the power distribution in the frequency and wavenumber domain, which is calculated by the 2-D Fourier transforming of fluctuating electric fields. The data set of fluctuating electric fields used in Figure 2(b) is selected over the space interval from $z = 220\rho_e$ to $500\rho_e$ (cold electron side), while the data used in Figure 2(c) is from $z = 50\rho_e$ to $150\rho_e$ (hot electron side). In Figure 2(b), $\omega_{pe}$ and $\lambda_{De}$ denote the electron plasma frequency and cold electron Debye length, respectively. The frequency of electric field in the cold-electron region is roughly located at $\omega = \omega_{pe}$, indicating that the fluctuating electric fields in the cold-electron region are Langmiur waves. As shown in Figure 2(c), the dominant wave mode in the hot-electron region mainly concentrates at $\omega = 0.32\omega_{pi}$ and $k = 0.45\lambda_{Dh}^{-1}$, where $\omega_{pi}$ and $\lambda_{Dh}$ are the proton plasma frequency and the hot electron Debye length, respectively. The wave number is consistent with that of the fastest growing linear mode ($k = 0.5\lambda_{Dh}^{-1}$) predicted from theoretical analysis (Manheimer 1977). The phase speed of the dominant wave mode is about $0.23v_{t0}$, which is approximately equal to the ion acoustic speed in the hot-electron region ($v_s = 0.26v_{t0}$). Thus, the parallel fluctuating electric fields in our simulations are Langmuir waves in the cold-electron region and ion acoustic waves in the hot-electron region. According to the previous simulation results in the uniform magnetic field (Li et al. 2012; Sun et al. 2019), Langmuir waves are excited by the hot electron streaming into the cold plasma and ion acoustic waves are generated by the return current. In addition, ion acoustic waves can evolve into a double layer, resulting in the suppression of electron heat flux.

Different from the results in uniform magnetic field, the whistle waves can be generated by the electron temperature anisotropy in the magnetic mirror field. We examine the temporal and spatial evolution of perpendicular fluctuating magnetic fields in Figure 3(a). To confirm that the excited fluctuations are whistle waves, we also diagnose the wave spectrum of the fluctuating magnetic fields in Figure 3(b). The wave spectrum is calculated by Fourier transforming of time series of fluctuating magnetic fields over the time interval from $\Omega_e t = 0$ to 300. Figure 3(c) shows the spatial distribution of fluctuating magnetic field at $\Omega_e t = 50$. The waves begin to be excited at about $\Omega_e t = 30$. Then, the waves propagate toward both sides parallel to background magnetic field. The position and time of the wave excitation coincide with the position and time of electron temperature anisotropy shown in Figure 1(c). The dominant wave mode mainly concentrates at $\omega = 0.27\Omega_e$. Since the wave length changes during the propagation in the non-uniform background magnetic field, we select one moment at linear growth stage to calculate the wave number in Figure 3(c). It is found that the wave length is about $20\rho_e$. Thus, the wave number is approximately $0.5\, \lambda_e^{-1}$, where $\lambda_e = c/\omega_{pe}$ is the electron inertial length. Using WHAMP model (https://github.com/irfu/whamp), we calculate the growth rate of the whistler waves based on the simulation parameter at $\Omega_e t = 50$ and find that the fastest growing linear mode mainly concentrates at $\omega = 0.3\Omega_e$ and $k = 0.5\, \lambda_e^{-1}$. The simulation result is consistent with the theoretical analysis. Therefore, the perpendicular fluctuating magnetic fields are the whistle waves excited by the electron temperature anisotropy.

To study the role of the whistler waves during the expansion of the hot electrons, we examine the electron distribution in the energy and pitch angle plane in Figure 4. The color in the figure denotes $N/N_{\text{tot}}$, where $N$ is the particle number in each $(E_e, \theta)$ bin, and $N_{\text{tot}}$ is the total particle number in the statistical interval. The statistical interval is $110 < z < 160$. The white line in Figure 4(c) and 4(d) represents the cyclotron resonance condition for whistler waves interacting with electrons. Since the wave length gradually changes during the propagation in the non-uniform background magnetic field, we select the minimum wave length ($\sim 20\rho_e$) and maximum wave length ($\sim 30\rho_e$) to calculate the cyclotron resonance condition in the $(E_e, \theta)$ plane. At $\Omega_e t = 10$, the electrons are mainly distributed in large pitch angle ($30° < \theta < 150°$). These large pitch angle electrons are confined by the magnetic mirror. At $\Omega_e t = 50$, the electron flux with the pitch angle near $135°$ shows obvious enhancements, which is attributed to the return current. The confinement of large pitch angle electrons can cause the electron temperature anisotropy. This leads to the excitation of the whistler waves which will interact with the electrons. At $\Omega_e t = 100$, the resonant electrons are scattered to small pitch angle by the whistler waves. Then, those particles can escape from the hot-electron region. At $\Omega_e t = 150$, more electrons are scattered to small pitch angle ($\theta < 30°$). Therefore, although the electrons with large pitch angle can be trapped by the magnetic mirror in the source region, the whistler waves will scatter these particles to small pitch angle, leading to the electrons escaping from the source region.

## IV. Conclusions and Discussion

In this paper, using a 1-D PIC simulation, we have addressed the expansion of hot electrons in a magnetic mirror field for the first time. The evolution of electron parallel temperature is similar to the simulation results in the uniform background magnetic field. The hot electrons with small pitch angle can freely expand into the cold plasma, which leads to the generation of Langmuir waves in the cold plasma. Ion acoustic waves can also be excited by the return current in the hot plasma. The evolution of perpendicular temperature is affected by the inhomogeneous magnetic field. The magnetic mirror is able to confine the electrons with large pitch angle, leading to the different evolution time scale between electron parallel and perpendicular temperature. This will cause the formation of electron temperature anisotropy, which can then generate the whistler waves near the interface between hot electrons and cold electrons. Although the large pitch angle electrons can be trapped by the magnetic mirror, the whistler waves will scatter these particles to smaller pitch angle through the cyclotron resonance. Therefore, the whistler waves may play an important role in the transport of electrons in the inhomogeneous magnetic field.

Previous simulations (Ishigura et al. 1985; Arber & Melnikov 2009; Sun et al. 2019) have primarily concentrated on the electron transport in a uniform magnetic field. Our research explores, for the first time, the effects of non-uniform magnetic field on the electron transport in solar flares using the PIC simulation. The formation of the thermal front in the non-uniform magnetic field is similar to that in the uniform magnetic field. Different from the results in uniform magnetic field, we have not observed obvious dissipation of the thermal front in the non-uniform magnetic field. The thermal front in the non-uniform magnetic field can persist longer time than that in the uniform magnetic field. In addition, the magnetic mirror is able to confine the large pitch angle electrons, which can lead to the generation of whistle waves. The previous studies (Melrose & Brown 1976) concluded that only electrons with $E_{||} > 100 keV$ can interact with whistler waves in the flaring magnetic loops and the whistler waves are not considered to scatter the electron which produce HXR emissions in those analysis. This is because they assumed that the frequency of whistler waves is $\omega \ll \Omega_e$. However, the whistler waves excited in our simulation is about $0.3\Omega_e$, and the energy threshold of resonant electrons can be reduced to about ten keV. Thus, the hot electrons which produce the HXR emissions may be scattered by the whistler waves in solar flares.

Here, we assume that the background electron temperature is 1 keV and the magnetic field is $10\ G$, which are typical parameters in coronal X-ray sources. Then, its electron gyroradius $\rho_e$ is about 0.1 m, and the length of our simulation box is about 400 m. Note that, to save computation source, the simulation box is designed to be much smaller than the realistic size of flaring loops which is about tens of thousands of kilometers. However, we think that the physical processes in our simulation may also occur in the flaring magnetic loops. In our simulation model, the magnetic mirror ratio in our concerned region where the main physical processes take place is about 2.5, which is a reasonable value in solar flares. As we all know that in the earth's radiation belt, the electron cyclotron frequency is about kHz, and the time scale of electron pitch angle scattering with whistle waves is approximately several hours (Ni et al. 2008). Based on this fact, we can infer reasonably that the time scale of electron pitch angle scattering with whistle waves in solar flares can reach a minute on the grounds that the electron cyclotron frequency is about MHz in solar flares. This time scale is comparable to the life time of observed HXR emission in corona. The time scale of Coulomb collision for tens of keV electrons is about a few hundred seconds in the solar corona (Krucker et al. 2008). Thus, the time scale of electron scattering by whistle waves is much smaller than that of the Coulomb collision. Our results indicate the whistler waves may play a key role in the transport of electrons which produce the HXR emissions in solar flares. It is interesting to note that the electron distribution may not be the Maxwellian distribution in the flaring magnetic loops. Numerical simulations of non-Maxwellian distribution electron transport in the inhomogeneous magnetic field will be performed in future works.

## Acknowledgements

This work was supported by the NSFC grant 41604128, 41631071, 41527804, 41774151, 41774169, Key Research Program of Frontier Sciences，CAS(QYZDJ-SSW-DQC010), Youth Innovation Promotion Association of Chinese Academy of Sciences (2016395), and Young Elite Scientists Sponsorship Program by CAST (2018QNRC001).

**Figure captions:**

**Figure 1.** The temporal and spatial evolution of (a) the parallel temperature of electrons, (b) the perpendicular temperature of electrons, and (c) the temperature anisotropy of electrons.

**Figure 2.** (a) The temporal and spatial evolution of the parallel fluctuating electric fields. (b) The dispersion relation of excited Langmuir waves. (c) The dispersion relation of excited ion acoustic waves.

**Figure 3.** (a) The temporal and spatial evolution of fluctuating magnetic fields $\delta B_x/B_0$. (b) The power spectrum of fluctuating magnetic fields $\delta B_{x\omega}^2/B_0^2$ obtained from the Fourier transform of $\delta B_x/B_0$. (c) The spatial profile of fluctuating magnetic fields $\delta B_x/B_0$ at $\Omega_e t = 50$.

**Figure 4.** The electron distribution in the energy and pitch angle plane at (a) $\Omega_e t = 10$, (b) $\Omega_e t = 50$, (c) $\Omega_e t = 100$, and (d) $\Omega_e t = 150$.

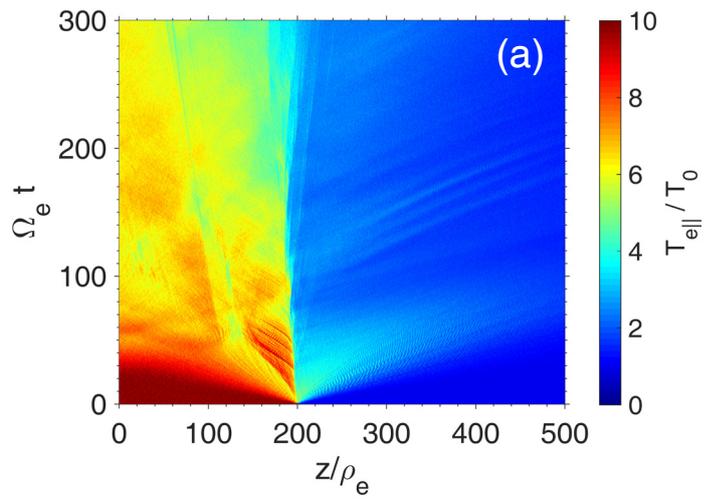
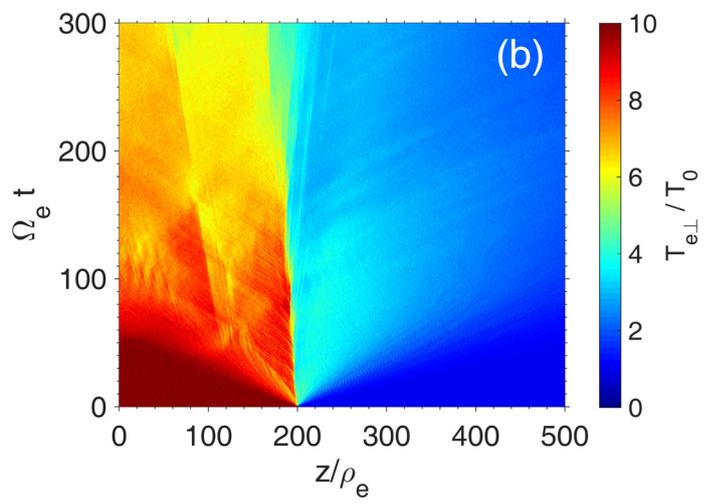
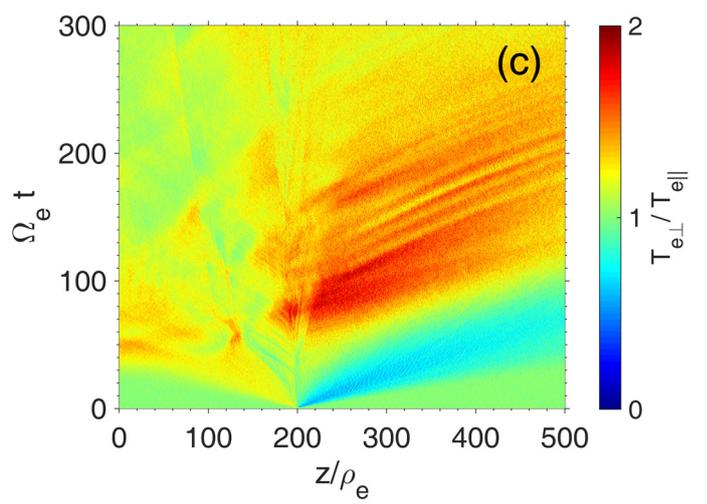

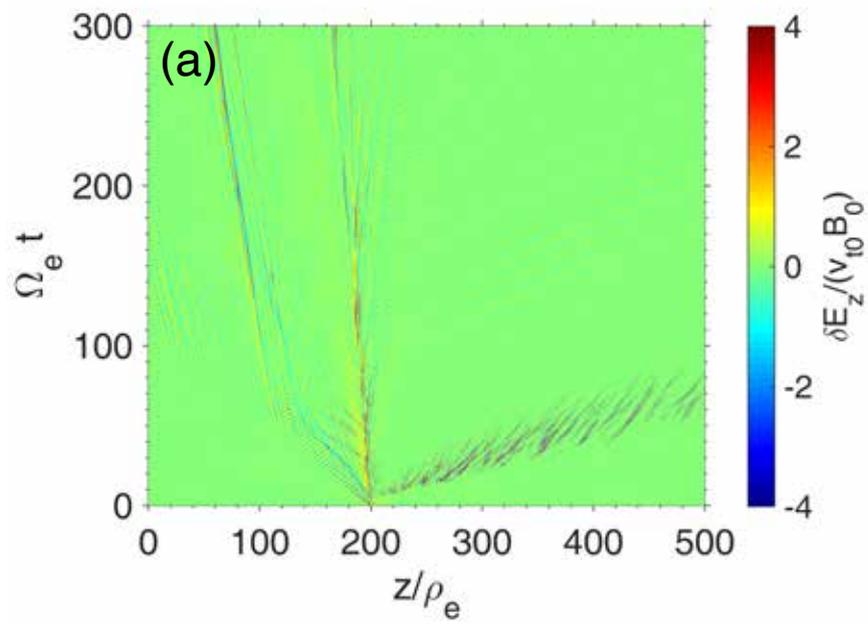
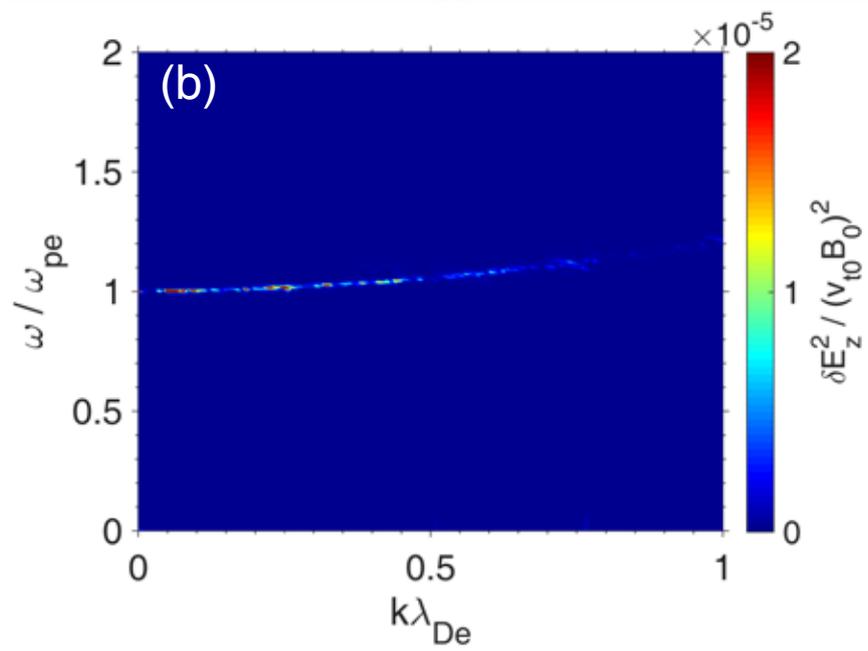
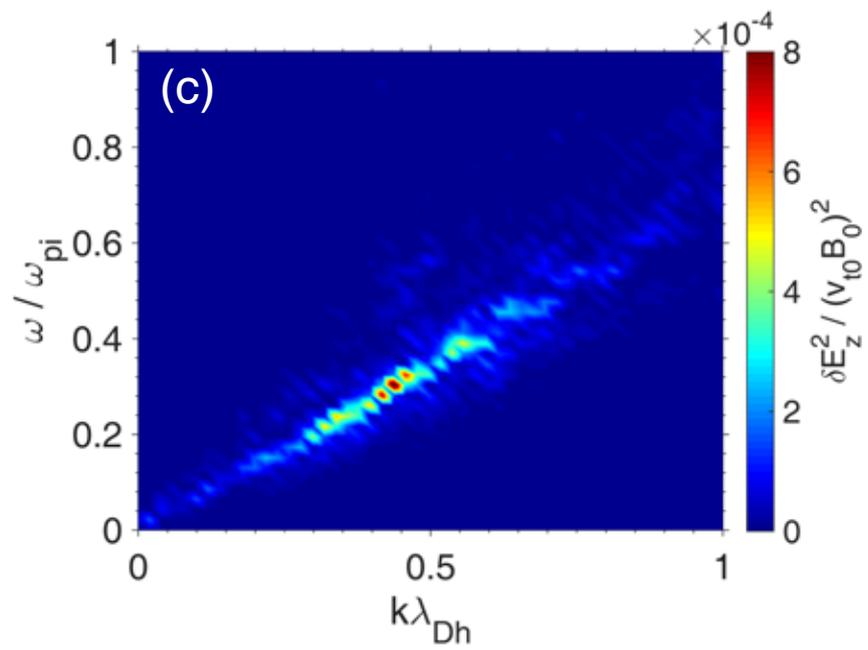

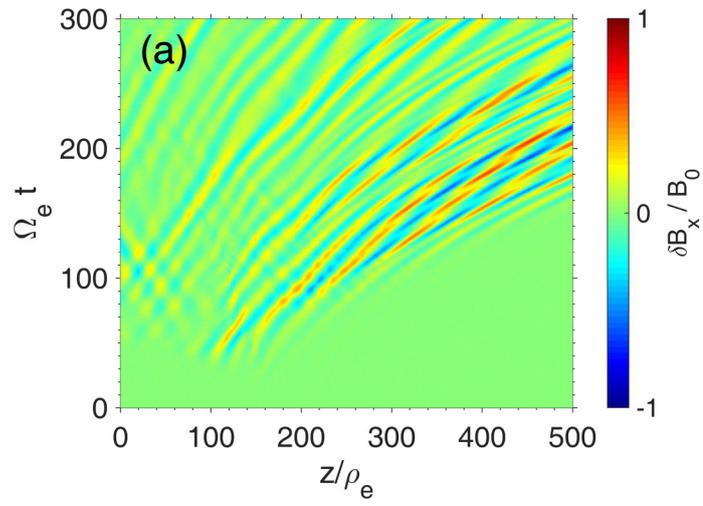

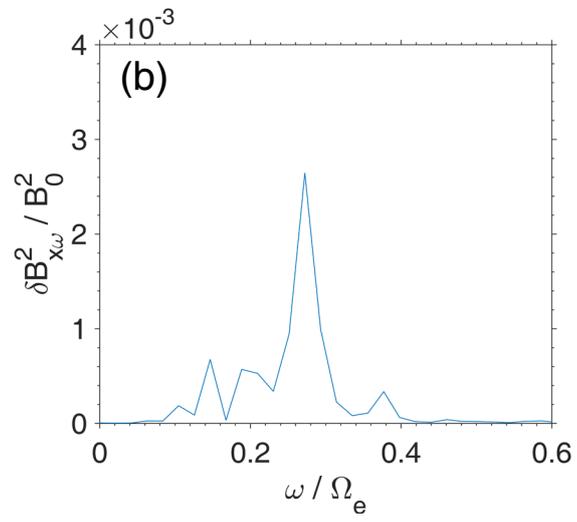

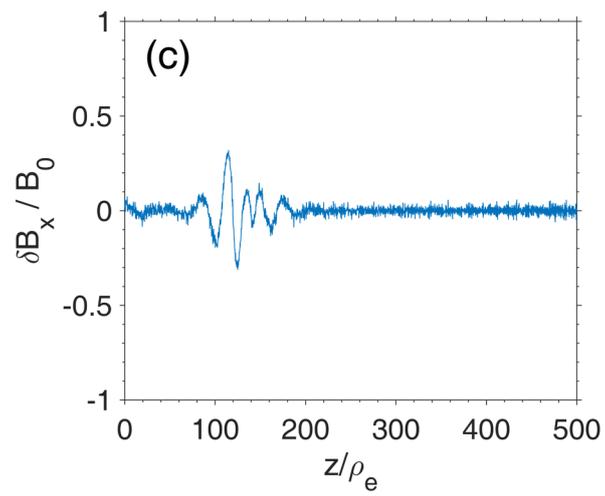

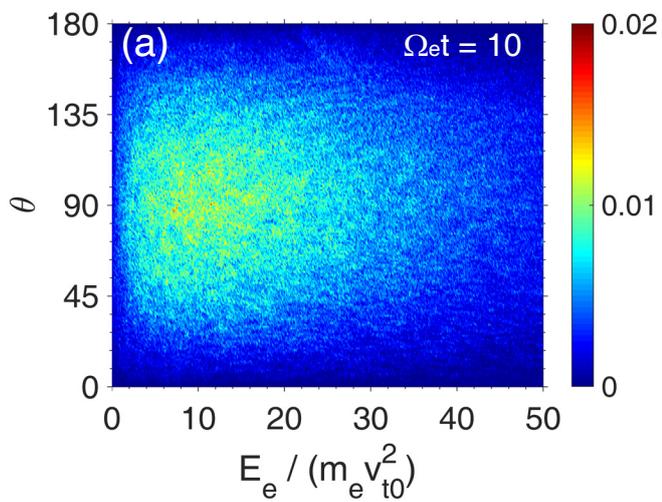
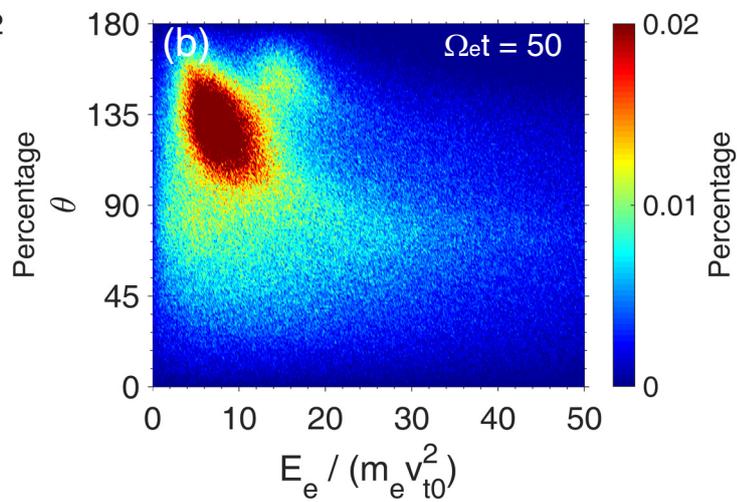
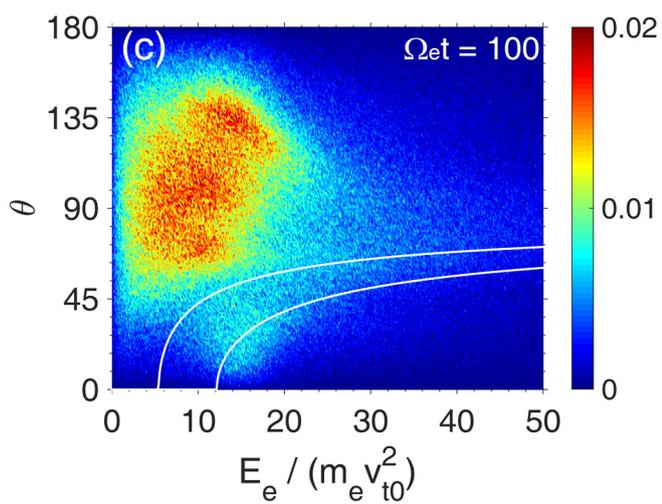
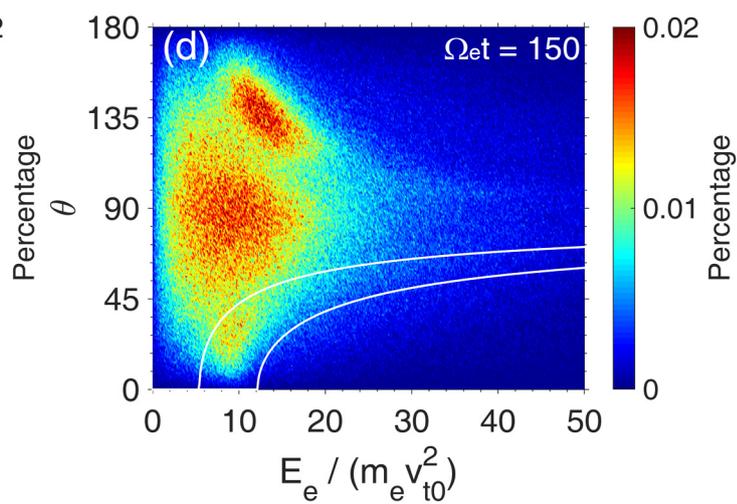